\documentclass[sn-mathphys,Numbered]{sn-jnl}% Math and Physical Sciences Reference Style

\usepackage{graphicx}%
\usepackage{multirow}%
\usepackage{amsmath,amssymb,amsfonts}%
\usepackage{amsthm}%
\usepackage{mathrsfs}%
\usepackage[title]{appendix}%
\usepackage{xcolor}%
\usepackage{textcomp}%
\usepackage{manyfoot}%
\usepackage{booktabs}%
\usepackage{algorithm}%
\usepackage{algorithmicx}%
\usepackage{algpseudocode}%
\usepackage{listings}%
\usepackage{upgreek}
\usepackage{siunitx}
\usepackage{blindtext}
\usepackage[rightcaption]{sidecap}

\begin{document}

\title[Magnetic microcalorimeters for primary activity standardization within the EMPIR project PrimA-LTD]{Magnetic microcalorimeters for primary activity standardization within the EMPIR project PrimA-LTD}

\author*[1]{\fnm{Michael} \sur{Müller}}\email{michael.mueller2@kit.edu}
\author[2]{\fnm{Matias} \sur{Rodrigues}}
\author[3]{\fnm{Jörn} \sur{Beyer}}
\author[2]{\fnm{Martin} \sur{Loidl}}
\author[1,4]{\fnm{Sebastian} \sur{Kempf}}

\affil[1]{\orgdiv{Institute of Micro- and Nanoelectronic Systems (IMS)}, \orgname{Karlsruhe Institute of Technology (KIT)}, \orgaddress{\street{Hertzstrasse 16}, \city{Karlsruhe}, \postcode{76187}, \country{Germany}}}
\affil[2]{\orgname{Université Paris-Saclay, CEA, List}, \orgdiv{Laboratoire National Henri Becquerel}, \orgaddress{\street{Bat. 602, CEA-Saclay}, \postcode{91120}, \city{Palaiseau}, \country{France}}}
\affil[3]{\orgname{Physikalisch-Technische Bundesanstalt (PTB)}, \orgaddress{\street{Bundesallee 100}, \city{Braunschweig}, \postcode{38116}, \country{Germany}}}
\affil[4]{\orgdiv{Institute for Data Processing and Electronics (IPE)}, \orgname{Karlsruhe Institute of Technology (KIT)}, \orgaddress{\street{Hermann-von-Helmholtz-Platz 1}, \city{Karlsruhe}, \postcode{76344}, \country{Germany}}}

\abstract{The precision of existing decay data of radionuclides for activity determination is often a limitation for actual applications in science, society, and industry. For this reason, the EMPIR project PrimA-LTD aims to introduce an advanced primary activity standardization technique that is based on magnetic microcalorimeters (MMCs) and that will offer very low energy threshold of few eV and a decay scheme independent detection efficiency close to $100\,\%$. As a proof of concept, we developed two MMC-based detector types in order to standardize an $\alpha$-decaying, a $\beta$-decaying and an electron capture decaying isotope. One detector type aims to introduce a reusable detector setup, while the other aims to provide highly accurate decay spectra by high resolution measurements with high statistics. We present the designs, fabrication status and first characterization measurements of both detectors types and outline next steps.}


\maketitle

\section{Introduction}
The precise knowledge of the activity and the decay spectrum of a given radioactive sample are important parameters in many fields of science, society, and industry: In nuclear medicine, precise decay data allows for an accurate calculation of the dose per administered activity, where uncertainties of activity standards in the order of $1\,\%$ are required \cite{judge2023NuclearMedicine}. The nuclear power industry uses decay data to determine the residual heat and its evolution over time in nuclear reactors and in nuclear waste management \cite{warwick2009NuclearWaste}. In radionuclide metrology, accurate decay data and activity determination techniques are needed for the definition of the Becquerel unit \cite{Debertin1996RadionuclideMetro, Pomme2022RadionuclideMetro}.

Liquid scintillation counting (LSC) \cite{kossert2015LSC}, as a widely established activity measurement technique, has a non-unity and energy dependent detection efficiency, and a comparably high energy threshold. Therefore, LSC exhibits dependencies on the decay-scheme and on the atomic and nuclear decay data. These aspects, depending on the nuclide to be measured, can lead to relative uncertainties of LSC-based activity measurements of a few per thousand \cite{Kossert2015Uncertainty}. As this limits the capabilities of the above mentioned applications, the EMPIR project\footnote{Link to the official PrimA-LTD homepage: \href{https://prima-ltd.net/}{https://prima-ltd.net/}.} ``PrimA-LTD - Towards new primary activity standardisation methods based on low-temperature detectors" aims to demonstrate a new primary decay scheme independent activity standardization method with near-unity quantum efficiency by using magnetic microcalorimeters (MMCs) \cite{fleischmann2005, kempf2018MMC} to achieve uncertainties as low as $0.1\,\%$. Comprising an ultra-sensitive paramagnetic temperature sensor placed in a weak magnetic field as well as an appropriate readout circuit, MMCs convert an energy input within a particle or radiation absorber into change of magnetic flux that can be sensed with utmost precision using state-of-the-art dc-SQUIDs \cite{dcSQUIDs2007}. Thanks to the resulting excellent energy resolution, MMCs have a very low energy threshold down to a few eV \cite{Krantz2023}. Furthermore, MMCs have demonstrated significant improvements of fundamental decay data determination by more than a factor of two, owing their superior energy resolution \cite{Kossert2022, Rodrigues2023}.

\section{Magnetic microcalorimeters for PrimA-LTD}
\begin{figure}[t]
    \centering
    \includegraphics[width=.95\linewidth]{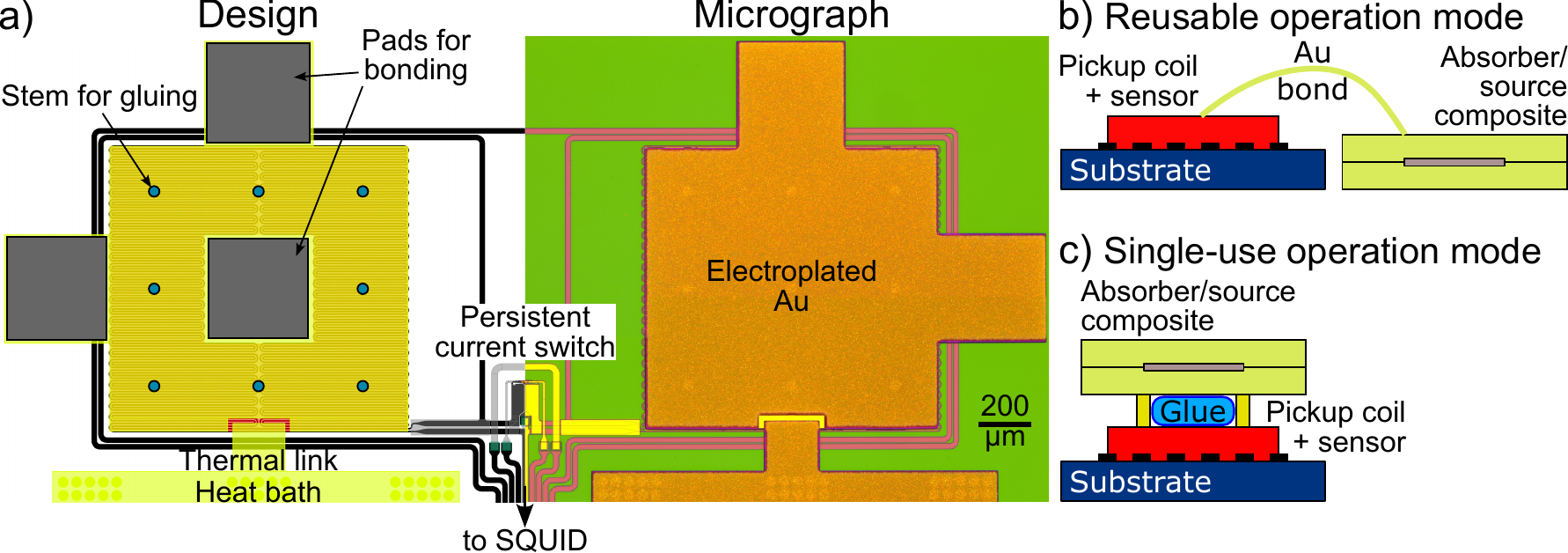}
    \caption{a) Overview of the RoS-L detector showing the design layout of the left pixel and a micrograph of the right pixel. The area covered by sensor material is indicated in yellow in the design layout. Moreover, b) and c) show the two methods for coupling of an absorber/source composite to the detector.}
    \label{fig:RoS}
\end{figure}

Within PrimA-LTD, we developed two novel detector types to measure the decay energy spectra (DES)\footnote{For a summary of DES microcalorimeter measurements refer to the summary paper \cite{Koehler2021DESwithLTDs} of Koehler et al.} of sources that are embedded into absorbers in a $4\pi$ geometry: the {\it RoS detectors}\footnote{Acronym for ``\textbf{R}eusable \textbf{o}r \textbf{S}ingle-use'' detectors.} for enhancing the flexibility of MMC-based measurements for radionuclide metrology, and the {\it implanted-$^{\textit{55}}$Fe detector} for high-resolution, high-statistics measurements of the $^{55}$Fe energy spectrum to determine fractional electron-capture probabilities. The RoS detectors (see Fig.~\ref{fig:RoS}) are designed for the usage of external absorber/source composites that are attached to the detector pixels \cite{Hadid2023}. One variant, named RoS-L, is optimized for measuring the DES of the $^{241}$Am alpha decay and the $^{129}$I beta decay, the other, named RoS-M, for measuring the electron capture decay of $^{55}$Fe. As the commonly used technique of gluing the external absorber to the detector pixel (see Fig.~\ref{fig:RoS}b) \cite{paulsen2019MMCGlue, loidl2020MetroBeta} makes an MMC usable only for a single experiment, the RoS detectors introduce an innovative design that allows to couple the absorber to the detector via wire bonds. This makes the detector reusable for multiple measurements and hence greatly improves flexibility and is resource-saving. To test and validate the performance of detectors with wire-bonded absorbers, the design enables a direct comparison of bonding and gluing by providing both, pads for bonding as well as stems for absorber gluing (see Fig.~\ref{fig:RoS}a).

The implanted-$^{55}$Fe detector is designed as a high resolution spectrometer for measuring fundamental decay data of $^{55}$Fe by providing the best achievable energy resolution for a detection efficiency of $>\SI{99.99}{\percent}$ over the full decay spectrum. Therefore, the detector concept allows for an integration of radionuclide into a microfabricated absorber. As drop deposition of the source typically leads to spectral distortions due to self-absorption \cite{le2012sourcecrystals} and electroplating of the source can only be used for selected metallic radionuclides, ion-implantation will be used to suppress any influence of the source on the measured spectrum. To prevent athermal phonon loss \cite{fleischmann2009, kempf2018MMC}, the microfabricated absorber is free-standing on stems and fully made of highly-conductive electroplated Au (residual resistivity ratio $> 30$), enabling fast thermalization and thus avoiding a dependence of the signal shape on the position of particle or radiation absorption within the absorber.

\section{Design and fabrication of the RoS detectors}
The RoS detectors are optimized to match external absorbers having a heat capacity of $\SI{400}{pJ/\K}$ (RoS-L) and $\SI{110}{pJ/\K}$ (RoS-M) at a temperature of $T=\SI{20}{mK}$. Both variants were designed to be fabricated on the same Si wafer. Each detector chip has a size of $\SI{4}{mm}\times \SI{5}{mm}$ and contains one two-pixel detector of one of the variants, that can be read out by a single dc-SQUID. In Fig.~\ref{fig:RoS}a), the RoS-L variant is shown as example. The pixel size is $\SI{1120}{\upmu m}\times \SI{1070}{\upmu m}$ with an area of the Ag:Er$\mathrm{_{475\,ppm}}$ sensor of about $\SI{1}{mm^{2}}$. The sensor is fully underlaid with a meandered Nb pickup coil having a line width of $\SI{5}{\upmu m}$ and a pitch of $\SI{10}{\upmu m}$. The Nb leads for injecting a magnetic field generating persistent current into the pickup coils are laid around the detector pixels in order to avoid vertical interconnections that may lead to reduced ampacity of the coil and thus could potentially limit the persistent current. The right pixel has a slightly reduced sensor area (reduced to \SI{95}{\percent}), to enable monitoring of heat bath temperature variations during operation. In the center of each pixel, an area of about $\SI{400}{\upmu m}\times \SI{400}{\upmu m}$ is free of sensor and is designated for wire-bonding of an external absorber, in case of the reusable operation mode. Two further bond pads are provided at the edges of the pixels and near the edge of the detector chip. The sensor and bond pad areas are covered with a $\SI{3}{\upmu m}$ thick layer of highly-conducting electroplated Au, allowing for homogeneous heat transfer from the bond pad to the sensor. Additionally, Au stems having a diameter of $\SI{20}{\upmu m}$ and a thickness of $\SI{3}{\upmu m}$ are electroplated in a second step, allowing for gluing an external absorber onto the detector pixel in the single-use mode. Both detector operation modes are schematically depicted in \ref{fig:RoS}b) and c). The RoS-M variant has a smaller pixel size of $\SI{490}{\upmu m}\times \SI{500}{\upmu m}$ with no bond pad in its center. However, the pixels are equipped with stems and bond pads at their edges.

\begin{SCfigure}
    %\centering
    \includegraphics[width=.64\linewidth]{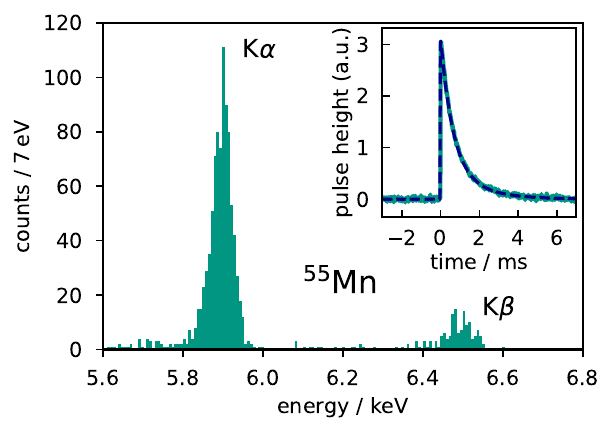}
    \caption{Measured spectrum taken during a first RoS-L characterization measurement showing the K$\alpha$ and K$\beta$ lines of an external $^{55}$Fe calibration source. Inset: Single (green line) and averaged (dashed blue line) detector signal. The detector was operated without particle absorber and at a cryostat temperature of \SI{15}{mK}.}
    \label{fig:RoS_daten}
\end{SCfigure}

The design values and the expected detector performance of the RoS detectors are summarized in Tab.~\ref{tab:Detector_properties}. From numerical simulations, taking into account the expected signal size based on the thermodynamic properties and geometry of sensor and absorber as well as all relevant noise contributions \cite{kempf2018MMC}, the energy resolution at \SI{20}{mK} is expected to be $\Delta E_{\mathrm{FWHM}}=\SI{38.5}{eV}$ and $\Delta E_{\mathrm{FWHM}}=\SI{19.5}{eV}$ for the RoS-L and the RoS-M variant, respectively. The energy resolution that is achieved in an experiment, however, will heavily depend on the thermal connection between absorber and sensor. In the reusable operation mode, longer signal rise times result from the bonding wires that form a thermal bottleneck between absorber and sensor. Furthermore, thermal connection between the external absorber and its substrate needs to be sufficiently prevented in order to avoid energy loss.

\begin{table}[t]
\centering
\begin{tabular}{@{}ll|ccc@{}}
\toprule
\multicolumn{2}{l|}{Detector type}                                                                                  & RoS-L                        & RoS-M                        & Implanted-$^{55}$Fe detector           \\ \midrule
\multicolumn{2}{l|}{Absorber heat capacity}                                                                         & $\SI{400}{pJ/K}$             & $\SI{110}{pJ/K}$             & $\SI{1}{pJ/K}$               \\
\multicolumn{1}{c|}{\multirow{3}{*}{\begin{tabular}[c]{@{}c@{}}SQUID\\ readout\end{tabular}}} & Input inductance    & $\SI{27}{nH}$                & $\SI{2}{nH}$                 & $\SI{2}{nH}$                 \\
\multicolumn{1}{c|}{}                                                                         & Mutual inductance   & $\SI{0.9}{nH}$               & $\SI{0.36}{nH}$              & $\SI{0.36}{nH}$              \\
\multicolumn{1}{c|}{}                                                                         & Current sensitivity & $\SI{2.3}{\upmu A/\Upphi_0}$ & $\SI{5.4}{\upmu A/\Upphi_0}$ & $\SI{5.4}{\upmu A/\Upphi_0}$ \\
\multicolumn{2}{l|}{Line width of pickup coil}                                                                      & $\SI{5}{\upmu m}$            & $\SI{5}{\upmu m}$            & $\SI{3}{\upmu m}$            \\
\multicolumn{2}{l|}{Pitch of pickup coil}                                                                           & $\SI{10}{\upmu m}$           & $\SI{10}{\upmu m}$           & $\SI{6}{\upmu m}$            \\
\multicolumn{2}{l|}{Single meander inductance}                                                                      & $\SI{28.5}{nH}$              & $\SI{6.79}{nH}$              & $\SI{0.92}{nH}$              \\
\multicolumn{2}{l|}{Sensor area (square shape)}                                                                     & $(\SI{1015}{\upmu m})^2$     & $(\SI{495}{\upmu m})^2$      & $(\SI{135}{\upmu m})^2$      \\
\multicolumn{2}{l|}{Sensor heat capacity}                                                                           & $\SI{297}{pJ/K}$             & $\SI{71}{pJ/K}$              & $\SI{1}{pJ/K}$               \\ \midrule
\multicolumn{2}{l|}{Expected energy resolution}                                                                     & $\SI{38.5}{eV}$              & $\SI{19.5}{eV}$              & $\SI{2.1}{eV}$               \\ \bottomrule
\end{tabular}
\caption{Design values and expected energy resolution at $T = \SI{20}{mK}$ of all detectors developed for PrimA-LTD. The detectors are designed to have a signal rise time of \SI{1}{\upmu s} and a signal decay time of \SI{1}{ms}. A SQUID white noise level of $\SI{0.15}{\upmu\Upphi_0/\sqrt{\mathrm{Hz}}}$ and SQUID $1/f$ noise level at \SI{1}{Hz} of $\SI{3}{\upmu\Upphi_0/\sqrt{\mathrm{Hz}}}$ is assumed.}
\label{tab:Detector_properties}
\end{table}

Once the fabrication of the RoS detectors was finished, the circuit functionalities were checked at low temperature by irradiating a bare RoS-L detector chip without connected absorbers with X-rays emitted by an external $^{55}$Fe calibration source. Fig.~\ref{fig:RoS_daten} shows a single and an average detector signal as well as a measured spectrum when operating the detector at a cryostat temperature of about $\SI{15}{mK}$. The pulse energies were determined after optimal filtering while rejecting pile-up events. The measured energy resolution of $\Delta E_{\mathrm{FWHM}}=\SI{58.8}{eV}$ was calculated by fitting the K$\alpha$ line with a single Gaussian. The measured energy resolution was expected to be higher than the predicted value because of an elevation of the MMC chip temperature due to heat dissipation on the adjacent SQUID readout chip that was operated in single-stage. Nevertheless, basic detector functionality was approved and can also be assumed for the RoS-M type, as both types were fabricated on the same wafer.

\section{Design and fabrication of the implanted-$^{55}$Fe detector}
\begin{figure}[t]
    \centering
    \includegraphics[width=\linewidth]{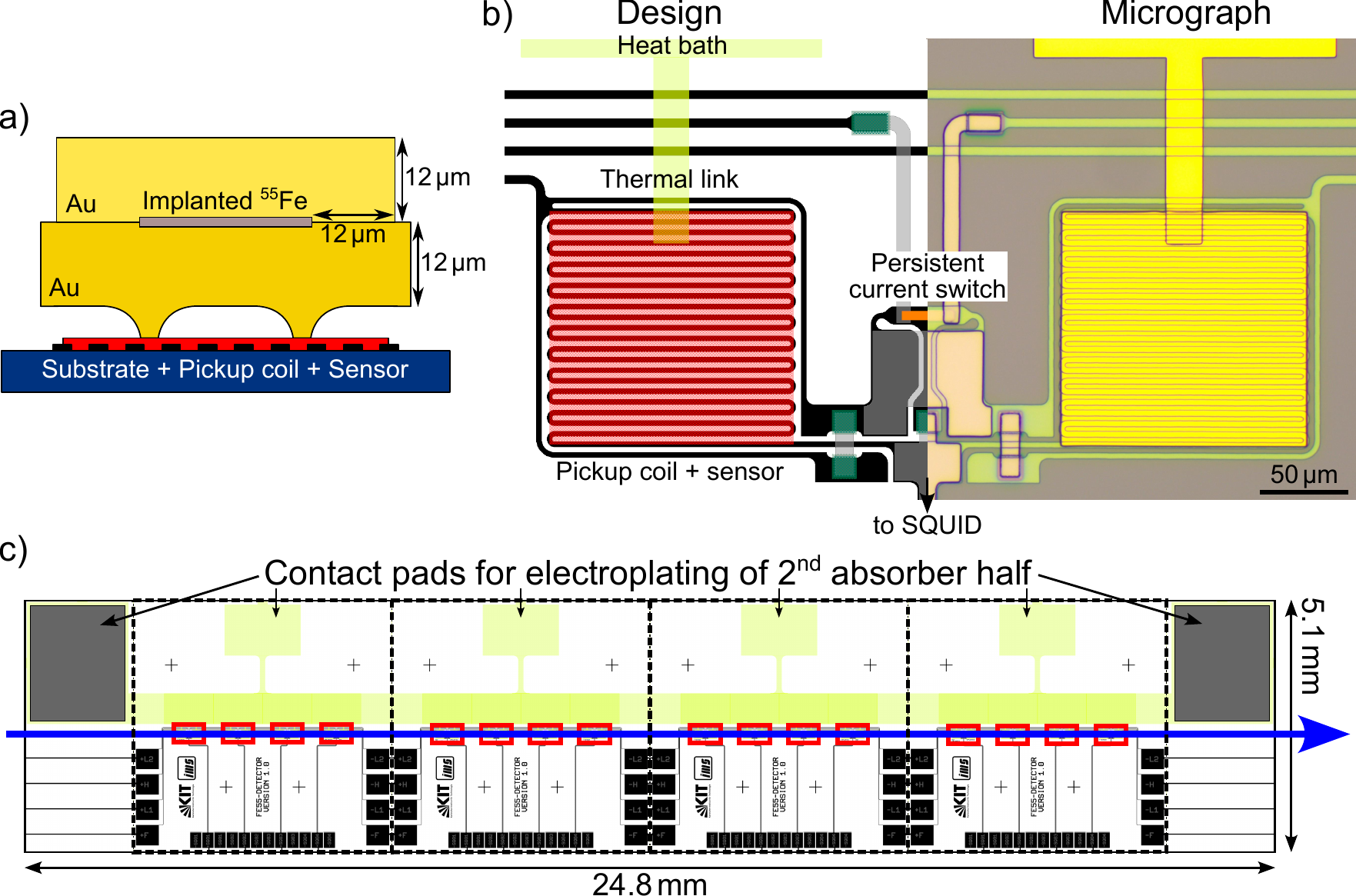}
    \caption{a) Schematic cross section of a microfabricated 4$\pi$ absorber geometry with ion-implanted $^{55}$Fe to be used with the implanted-$^{55}$Fe detector with a detection efficiency of $>\SI{99.99}{\percent}$. The absorber is free-standing on stems to prevent athermal phonon loss. Both absorber halves are electroplated by means of a special electroplating process. b) Overview of the implanted-$^{55}$Fe detector showing the design layout of the left and a micrograph of the right pixel up to the sensor layer. c) Design layout of a full chip to be ion-implanted with $^{55}$Fe. It contains a total of 16 individual detectors with 32 pixels, 20 of which are going to be implanted. The positions of the detectors are indicated by red boxes, each containing one pixel pair as shown in b). The ion-implantation path is indicated by a blue arrow. The contact pads on the very left and right are used for electroplating of the 2$^{\mathrm{nd}}$ absorber half. After completed fabrication, the chip can be further diced in up to 4 single chips (indicated by dashed black boxes).}
    \label{fig:Fe55}
\end{figure}

To detect the low energy photons ($E<\SI{6.5}{keV}$) emitted by the electron capture decay of $^{55}$Fe with a detection efficiency of $>\SI{99.99}{\percent}$, a $4\pi$ enclosure of the ion-implanted $^{55}$Fe by absorber material is necessary (see Fig.~\ref{fig:Fe55}a). A Au thickness of \SI{12}{\upmu m} was estimated by performing Monte-Carlo simulations using the code PENELOPE 2018 \cite{PENELOPE2018}. With an absorber size of about $\SI{170}{\upmu m}\times \SI{170}{\upmu m}$, an area of $\SI{136}{\upmu m}\times \SI{136}{\upmu m}$ is designated to be ion-implanted. The total heat capacity of the absorber amounts to \SI{1}{pJ/K} at $T=\SI{20}{mK}$. The implanted-$^{55}$Fe detector is optimized specifically for this absorber geometry, having a double-pixel layout that can be read out by a single dc-SQUID, similarly to the RoS detectors. The meander-shaped Nb pickup coils have a line width of \SI{3}{\upmu m} and a pitch of \SI{6}{\upmu m}. Fig.~\ref{fig:Fe55}b) shows different views of the implanted-$^{55}$Fe detector. The integration of the $4\pi$ enclosure of $^{55}$Fe is performed as follows: (i) electroplating of the 1$^{\mathrm{st}}$ absorber half, (ii) masking the absorber region designated for ion-implantation, (iii) ion-implantation of $^{55}$Fe and (iv) electroplating of the 2$^{\mathrm{nd}}$ absorber half. All detector pixels are fabricated up to the 1$^{\mathrm{st}}$ absorber half on wafer scale, then the wafer is diced into separate chips as depicted in Fig.~\ref{fig:Fe55}c), as the space in the ion-implantation facility \cite{kieck2019RISIKO} is limited. On a chip, the detector pixels are arranged in a straight line, allowing for a straight ion-implantation path. During ion-implantation, the whole chip, except for the absorber areas, is covered with photoresist to avoid parasitic implantation at the chip edges. Electroplating of the 2$^{\mathrm{nd}}$ absorber half is performed by using the outer two contact pads that connect the electroplating setup to the heat bath. The heat bath itself is in galvanic connection to the area to be electroplated via the thermal links and sensor layers of the the individual pixels. After fabrication, the chips can either be used as they are or be further diced in up to four single chips. Each single chip contains a total of eight pixels, five being ion-implanted, two being dedicated for temperature monitoring and one for measuring the background spectrum. Furthermore, each single chip is equipped with an extra contact pad in case a further electroplating step is needed.

As the fabrication of the \SI{24}{\upmu m} thick, fully electroplated and stacked particle absorbers requires careful optimization of photoresist processing and electroplating parameters, we developed a specific absorber fabrication process for the implanted-$^{55}$Fe detectors. This process is described in detail in an upcoming publication. Basic detector functionality was successfully demonstrated by irradiation with an external $^{\mathrm{83m}}$Kr source. The implanted-$^{55}$Fe detector pixels used for this measurement were fabricated up to the 1$^{\mathrm{st}}$ absorber half and were not implanted with $^{55}$Fe. Determination of the achievable energy resolution has not been possible yet due to low statistics and unstable operation temperature. From numerical calculations, performed similarly as for the RoS detectors, the predicted energy resolution at \SI{20}{mK} is $\Delta E_{\mathrm{FWHM}}=\SI{2.1}{eV}$. The design values of the implanted-$^{55}$Fe detector are summarized in Tab.~\ref{tab:Detector_properties}.

\section{Conclusion}
Magnetic microcalorimeters (MMCs) are an ideal detector technology for determining decay data of radionuclides thanks to their excellent energy resolution and high detection efficiency. For this reason, the EMPIR project PrimA-LTD aims to demonstrate MMC-spectrometry as a novel method for primary activity standardization. As a proof of concept, we developed two types of MMC-based detectors that enhance the flexibility and accuracy of decay measurements within radionuclide metrology. The RoS detectors will improve the flexibility and throughput of activity measurements by introducing a reusable detector setup. The implanted-$^{55}$Fe detector will improve the accuracy of decay spectra by introducing ion-implantation into microfabricated absorbers as a new source preparation technique in the context of radionuclide metrology. Basic functionality of both detector types has been verified by first evaluation measurements using X-rays from external $^{55}$Fe and $^{\mathrm{83m}}$Kr sources and the 2-step electroplating of the microfabricated absorbers for the implanted-$^{55}$Fe detector has been successfully demonstrated. Next steps are the comparison of the RoS detector performances between glued and wire bonded absorber foils as well as the completion of the implanted-$^{55}$Fe detector fabrication after ion-implantation.

\bmhead{Acknowledgments}
This project 20FUN04 PrimA-LTD has received funding from the EMPIR programme co-financed by the Participating States and from the European Union’s Horizon 2020 research and innovation programme. Moreover, we thank Andreas Reifenberger from the Kirchhoff-Institute for Physics at Heidelberg University for the fabrication of the RoS detector wafer. In addition, M. Müller greatly acknowledges financial support by the Karlsruhe School of Elementary Particle and Astroparticle Physics: Science and Technology (KSETA). 

\bibliography{bibliography.bib}

\end{document}